 \documentstyle[twoside,fleqn,espcrc2,epsfig]{article}

\newcommand{\m}{\medbreak}
\newcommand{\no}{\noindent}
\newcommand{\EQ}{\begin{equation}}
\newcommand{\eq}{\end{equation}}
\newcommand{\EQA}{\begin{eqnarray}}
\newcommand{\eqa}{\end{eqnarray}}

\newcommand{\ET}{\mbox{$E_T\ $}}

\newcommand{\AL}{\mbox{$A_{L}\ $}}

\def\prl#1#2#3{ Phys. Rev. Lett. {\bf{#1}}, #2 (#3)}

\def\epj#1#2#3{ Eur. Phys. J. {\bf{#1}}, #2 (#3)}

\newcommand{\AmS}{{\protect\the\textfont2
  A\kern-.1667em\lower.5ex\hbox{M}\kern-.125emS}}

\title{New Physics at Polarized Hadronic Colliders}

\author{J.-M. Virey\address[cptup]{Centre de Physique 
Th\'eorique (UPR 7061), CNRS-Luminy, Case 907,
F-13288 Marseille Cedex 9, France\\ 
and \\ 
Universit\'e de Provence, Marseille, France}%
\thanks{Preprint CPT-01/P.4243. To appear in the proceedings
of the ECT* workshop ``The Spin Structure of the Proton 
and Polarized Collider Physics'', 
July 2001, Trento, Italy.}
   \hspace*{0.1cm}     and
        P. Taxil\addressmark[cptup]}
       
\begin{document}

\begin{abstract}
Concerning new physics beyond the Standard Model
we explore the discovery and analysis potentials of polarized (hadronic) 
experiments and we compare with the
 unpolarized case. For discovery, beam polarization is helpful
in the case of purely hadronic new interactions. In any case, beam polarization
provides us a unique piece of information on the chiral and flavour structures.
\vspace{1pc}
\end{abstract}

\maketitle

\section{Introduction}

At the Brookhaven National Laboratory it is expected, during the year 2001,
to run the Relativistic Heavy Ion Collider (RHIC) in the polarized
$\vec p \vec p$ mode.
Actually, this run will be done at an energy  $\sqrt s = 200 $ GeV and
with a luminosity of a few $10^{30} cm^{-2}s^{-1}$.
The nominal energy  $\sqrt s = 500 $ GeV and  luminosity
${\cal L}\ =\, 2. 10^{32} cm^{-2}s^{-1}$ should be reached
in the early months of 2003, allowing an exposure
of 800 $pb^{-1}$ in four months running.\\

The physics potential of the RHIC-Spin experiments
has been extensively covered in a recent review paper \cite{BSSW}.
In view of the apparent feasability of accelerating polarized protons
at very high energies, several other projects are
now considered. 
Indeed, since 1997 several workshops have been organised
on the physics interests of the HERA machine with a beam of polarized
protons \cite{HERAw}. More recently, the future of the LHC
has been considered by a working group at CERN. It appears that a polarized
option for the proton beams is not excluded. In this workshop, Albert de Roeck
reports on such a future polarized LHC option \cite{deRtrento}.
Note that the physics with polarized beams at hadronic supercolliders
has been already explored some time ago \cite{Bourrely,Penn}.\\

The physics program of the RHIC and HERA polarized experiments is directly connected
to the analysis of the QCD structure of the proton.
For instance, at RHIC the first part of the program will 
include precise measurements
of the polarization of the gluons, quarks and sea-antiquarks
in a polarized proton. This will be done thanks to well-known
Standard Model (SM) processes : direct photon, $W$ and $Z$ production, 
Drell-Yan pair production, heavy-flavor production
and the production of jets.
The helicity structure of perturbative QCD will be thoroughly
tested at the same time with the help of Parity Conserving
double spin asymmetries.\\

However, here we ask an other question:``Concerning New Physics,
which kind of information can be obtained from such polarized
facilities ?''

Of course, the answer is far from trivial. It depends on the various
properties of the machine under consideration, like the energy scale,
luminosity, type of incident particles, type of process which is analysed ...
Nevertheless, several definite advantages of the polarized experiments
can be pointed out simply by comparison with the existing results of
unpolarized experiments.\\

In the following we will consider some possible manifestations of 
new physics models which are
as most as possible model independent. In particular we consider the
presence of Contact Interactions (CI) which mimic the existence
of new subconstituants or new interactions. We also consider the presence
of new gauge bosons ({\it e.g.} $Z'$, $W'$ or Leptoquarks). In this case
one can get a resonant behaviour if the energy is sufficiently high. 
Conversely we do not consider for the moment the SUSY phenomenology 
which is certainly very rich but highly model dependent.\\

Presently, the strongest constraints on such new physics models 
are obtained from LEP, HERA and Tevatron colliders. In particular,
for {\it\bf processes involving leptons}, either in the initial state
like at LEP and HERA, or in the final state like the Drell-Yan process
at Tevatron, some strong and complementary constraints
have been obtained on new physics models.
 For instance, the characteristic energy scale $\Lambda$
of CI is constrained to be above $4\, TeV$ for lepton-lepton or
lepton-quark CI \cite{PDG}. 
The mass of a $Z'$ which is present in popular GUT models (extra $U(1)$'s
or Left-Right models) 
should be above $600-700\, GeV$ \cite{PDG}. The precise values of
the bounds depend on the details of the models for each collider.

If polarization of the beams is present, we do not expect any improvements
on these constraints. Specific examples have been considered in
\cite{JMVHERACI} for $eq$-CI and in \cite{JMVHERALQ} for Leptoquarks,
in the context of HERA. It was shown that the polarization of the beams
is useless to reveal some new physics effects, but it is very useful to identify the
origin of the new interaction, if the latter is discovered first by 
the unpolarized  experiments.\\

Conversely, when we consider some {\it\bf pure hadronic processes} 
the situation is 
drastically different. Indeed, in a context like the
one of the Tevatron, it appears that
the analysis of hadronic processes, like jet or $b\bar b$ productions, is
far to be easy. In such channels the systematic errors, coming both
from the experimental and the theoretical sides, are rather large.
For instance, for jets the CDF and D0 collaborations are still
in disagreement \cite{D0jet,CDFjets2}. For $b\bar b$ production, the comparison
between data and theory is still problematic \cite{bbb}.
Therefore, the bounds on new physics models from such hadronic channels are
quite low in comparison to the ones from leptonic processes.
For example, the last published analysis of D0 on $qq$-CI gives a
limit of $\Lambda > 2.2\, TeV$ (from jet production) \cite{D0jet}, 
a factor two below the
constraints obtained on $eq$-CI from the same experiment (from the
Drell-Yan process).

In this case, the availability of polarized proton beams can be very helpful,
especially in the hope of a discovery. Indeed, the main difference between an 
unpolarized and a polarized experiment concerns the basic observables
used in the analysis of data : we shift from unpolarized cross sections
to spin asymmetries. The spin asymmetries have the net advantage
to minimize systematic errors in comparison to cross sections
\cite{BSSW}. This is especially true for channels with
huge systematics like pure hadronic processes.
For example, it has been shown that the RHIC Spin experiment is
competitive with the Tevatron
for the search of a $qq$-CI \cite{PTJMVCI} or of a leptophobic $Z'$ \cite{PTJMVZ'},
in spite of an energy four times lower !\\

To summarize we can say that polarized beams allow to define a
new type of observables : spin asymmetries.\\
Firstly, for processes with high systematic errors, like pure hadronic channels,
the use of spin asymmetries reduce strongly these systematics and
allow the polarized experiment to have a high discovery potential. In the next
section we examplify this result with the analysis of the effects of $qq$-CI and
leptophobic $Z'$ at RHIC (and at a polarized LHC in section 3).\\
On the other hand, the analysis of spin asymmetries yields a unique 
opportunity to get some informations on the chiral structure and also on the
flavour structure of the new interaction. 
Some examples in the context of HERA can be found
in \cite{JMVHERACI,JMVHERALQ}. In the third section
we show the results of an analysis done in the context of the LHC
with polarized protons, to pin
down the origin of a new generic $Z'$ boson from the Drell-Yan process
\cite{AFPT}.
\section{The RHIC case (jet production)}

\subsection{Observable and models}

At RHIC, running in the $\vec p \vec p$ mode, it will be possible to measure 
with a
great precision  the single Parity Violating (PV) asymmetry \AL :
\EQ
A_L \; =\; {d\sigma_{(-)}-d\sigma_{(+)}\over 
d\sigma_{(-)}+d\sigma_{(+)}}
\eq
\no where only one of the proton is polarized
(the sign  $\pm$ refer to the helicity of the polarized
proton). The cross section $d\sigma_{(\lambda)}$ means the one-jet
production cross section in a given helicity configuration, 
$p_1^{(\lambda)}p_2 \rightarrow jet + X$, estimated at  
$\sqrt{s}=500\, GeV$
for a given jet transverse energy \ET, integrated over a pseudorapidity interval 
$\Delta \eta =1$ centered at $\eta\,=\,0$.  \\

The new physics models that can be tested at RHIC with jet production are
the followings :\\

\no $\bullet$ First \cite{PTJMVCI} one can think to a 
simple phenomenological contact interaction
which could represent the consequences of quark compositeness. Such 
(color singlet and isoscalar) terms are
usually parametrized following Eichten et al. \cite{EHLQ} :

\EQ\label{Lcontact}
{\cal L}_{qq} = \epsilon \, {\pi\over {2 \Lambda^2}} 
\, \bar \Psi \gamma_\mu (1 - \eta \gamma_5) \Psi . \bar \Psi
\gamma^\mu (1 - \eta \gamma_5) \Psi
\eq
\noindent
where $\Psi$ is a quark doublet, $\epsilon$ is a sign, $\eta$ 
can take the values $\pm 1$ or 0 and $\Lambda$ is
the compositeness scale.
In the following we will consider the $LL^-$ case with Left-handed
chiralities ($\eta = 1$) and constructive interference with
QCD amplitudes which corresponds to $\epsilon = -1$.\\

\no $\bullet$ Second, we can consider some new neutral gauge bosons
with general Left and Right-handed couplings 
to each given quark flavor $q$:
\EQ
\label{lag}
{\cal L}_{Z'} = \kappa {g\over 2 \cos \theta_W} Z'^{\mu}{\bar q} \gamma_\mu[ C^q_{L}
(1 - \gamma_5) \; +\; C^q_{R} (1 + \gamma_5) ] q
\eq \no
the parameter $\kappa = g_{Z'}/g_Z$ being
of order one. 
A particular class
of models, called leptophobic $Z'$, is poorly constrained
by present data since they evade the constraints
coming from leptonic channels. 
Such models appear in several string-inspired scenarios
\cite{stringZ'}. Non supersymmetric models can also
be constructed \cite{NoSusy}. Here, for simplicity, we consider
only the model $A$ of \cite{PTJMVZ'} and we refer the reader
to this reference \cite{PTJMVZ'} for more details.
In addition, it was advocated in \cite{Cvetic} that such
a boson could appear with a mass close to the electroweak scale 
and a mixing angle to the
standard $Z$ close to zero.

\subsection{Results}

We give in Table 1 the 95 \% C.L. limits on $\Lambda \equiv \Lambda_{LL^-}$
(eq.2) one gets \cite{PTJMVup}, at lowest order, from a comparison between the 
SM asymmetry $A_L$ and the Non-Standard one. A $\chi^2$ analysis is used.\\

\begin{center}
\begin{tabular}{ccccc}
\hline
$L\; (fb^{-1})$ & 0.8 & 4 & 20 & 100 \\
$\Lambda\; (TeV)$ & 3.2 & 4.55 & 6.15 & 7.55 \\
\hline
\end{tabular} 
\end{center}
\begin{center}
Table 1 : Limits on $\Lambda_{LL^-}$ at 95\% CL
for RHIC with $\sqrt s = 500\, GeV$.\\
\end{center}

One can compare the bounds at $\sqrt s = 500$ GeV and $L = 0.8 fb^{-1}$ with
the ones after some luminosity upgrades : $L = 0.8 fb^{-1}$ is the nominal
luminosity expected after 4 months of run. $L = 20 fb^{-1}$ may be reached in
the future with the same amount of running time but after an upgrade
of the machine \cite{Saitospin}.
$4 fb^{-1}$ 
($100 fb^{-1}$) represents
5$\times\,$4 months running with the designed (future) nominal luminosity.\\

This table can be compared with the last published analysis
  of the D0 experiments at Tevatron \cite{D0jet}: 
$\Lambda > 2.2$ TeV (95\% C.L.) from the dijet mass cross section.
From these figures we have extrapolated
a limit at Tevatron of 3.2 TeV (3.7 TeV) with a 1 $fb^{-1}$ (10 $fb^{-1}$) exposure.
If we compare these numbers with the ones from Table 1, it is clear
that RHIC has a better sensitivity than Tevatron, if sufficient data
are accumulated.\\

Turning now to the case of a leptophobic $Z'$, we present in
Fig.1 the constraints on the parameter space ($\kappa, M_{Z'}$) obtained \cite{PTJMVup}
from $A_L$ in the flipped SU(5) model (model $A$ of \cite{PTJMVZ'}). 
The dotted curves correspond to ${\sqrt s} = 500$ GeV and the
dashed curves to ${\sqrt s} = 650$ GeV. From bottom to top
they correspond to an integrated luminosity
$L = 1,10,100 fb^{-1}$. It appears that the increase in luminosity
is more efficient than the increase in energy. 
Therefore the high
luminosity scenario has to be supported even if the RHIC $pp$
c.m. energy remains at its "low" value.\\

We display also in Fig.1 the inferred constraints coming 
from the published results of UA2 \cite{UA2}, CDF \cite{CDFjets2}
and D0 \cite{D0jets} experiments. The form of the forbidden areas
result from a combination of statistical and systematic errors.
For high $M_{Z'}$ one is looking for some unexpected high-$E_T$
jet events and the main uncertainty is statistical in nature.
For instance, the upper part of the "CDF area" is well below
the one of D0 because of the well-known excess observed by CDF at 
high-$E_T$. In the future (run II) the increase in statistics
will improve the bounds in the  ($\kappa, M_{Z'}$) plane by 
enlarging the upper part of the CDF and D0 areas (or will lead to 
a discovery). For relatively low $M_{Z'}$ values, the main problem
comes from the large systematic errors for "low" $E_T$ jets.
Due to these systematics, at Tevatron, even with a high statistics
it will be difficult
to probe the low $\kappa$ region for $M_{Z'} \leq 400$ GeV
or to close the windows around $M_{Z'} \simeq 300$ and 100 GeV. 
In this respect, as can be seen
from Fig.1, the RHIC-Spin measurements at high luminosity should
allow to cover this region and to get definite conclusions, if 
the new interaction violates parity.

\begin{figure}[h]
\centerline{\psfig{file={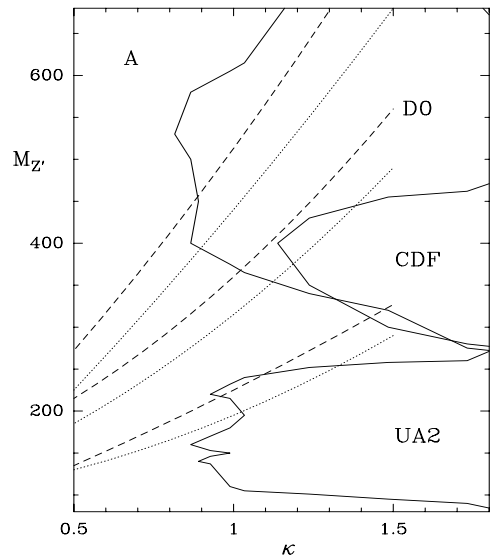},width=8truecm,height=10truecm}}
\vspace{-1.3cm}
\caption{Bounds on the parameter space for leptophobic 
flipped SU(5) $Z'$ models (see text).}
    \label{fig1} 
\end{figure}

To conclude, we have seen that RHIC is competitive/complementary
with the Tevatron to discover a new weak force belonging purely
to the quark sector.

\section{The LHC case}

The LHC will give acess to a completely new energy domain. Hence
the whole community is expecting to find, at least, some physics
beyond the SM. However, besides the problem of discovery, the problem
of the identification of the nature of a new interaction is fundamental
and much more difficult.

\subsection{Leptonic channels}

The existence of new vector bosons
is a common prediction of many scenarios going beyond the SM.
The most popular models are based on an expanded gauge
symmetry group which contains extra $U(1)$ and/or extra $SU(2)$
group(s) leading, after breaking of the symmetry, to the presence
of new massive gauge bosons. In general, the masses of these new
particles are not fixed by the theory. On the other hand their
couplings to ordinary fermions and gauge bosons are determined in
each particular model. Here we will not review all the existing models
but simply quote the models considered in \cite{AFPT}.
Namely, we consider the continuous $E_6$ , Left-Right (LR) and Y models.
We have simulated also more exotic models where the couplings
to fermions are completely fixed : the "Sequential
Standard Model" (SSM), the "BESS" models, and a preonic model labeled
by  "HYP". Additional models and all references can be found in \cite{AFPT}.\\

Concerning $Z'$ production, all the LHC studies
have shown that it will not be difficult to produce a massive $Z'$
and to detect it in the leptonic mode through the
Drell-Yan process. The discovery limit is around $M_{Z'}\simeq 5\, TeV$
from the analysis of the unpolarized cross section \cite{Atlas}. 
The sensitivities of the spin asymmetries are well below, roughly of the order
of $2\, TeV$ \cite{AFPT}. So, we recover the result that polarized beams
are not necessary {\it to discover} some new physics effects through
leptonic signatures.\\
Concerning the question of the identification, it has been explored by
many authors within unpolarized 
$pp$ collisions (see e.g. \cite{cvetichep93} and references therein).
The observables at hand are essentially
the forward-backward angular asymmetry $A_{FB}$ and some decay widths.
It turns out that the observables should allow to pin down
some particular models but not to treat some more general cases.
Then polarization should help.\\

Considering that only {\sl one} proton beam is polarized one can
focus on the process $\vec p_ap_b\rightarrow \ell^+\ell^-X$ in the
vicinity of the $Z'$ peak. 
Then, one can define two PV spin-dependent quantities :  
\EQ 
A_{LR}(Y,M)={d\sigma^{-}-d\sigma^{+}\over
d\sigma^{-}+d\sigma^{+}}
\eq
\EQ
A^{Pol}_{FB}(Y,M)={\left(d\sigma^{-}_F-d\sigma_B^{-}\right)-
\left(d\sigma^{+}_F-d\sigma_B^{+}\right)\over \left(d\sigma^{-}
_F+d
\sigma_B^{-}\right)+\left(d\sigma^{+}_F+d\sigma_B^{+}\right)}
\eq
\m
\noindent
Here $d\sigma^{\pm}$ stands for $d\sigma^{\pm}/dM$, 
$\pm$ corresponding to the helicity of the polarized
proton, $M$ being
the invariant mass of the lepton pair, and the cross sections are
integrated over some rapidity interval $\pm Y$. Note that
the measurement of the single spin asymmetry $A_{LR}$ does not
require the identification of the charge of the outgoing leptons.
$A^{Pol}_{FB}$ is a polarized forward-backward asymmetry which
reflects the angular dependence of the subprocess spin asymmetry.\\
It appears that $A_{LR}$ is independent
of the final state couplings  (leptonic or whatever) and that
$A^{Pol}_{FB}$ goes like the product of initial state and final
state couplings (as goes the well-known unpolarized $A_{FB}$).
This shows the complementarity of these two spin asymmetries.\\

We
present in Figs.2-3 the results of our calculations \cite{AFPT} under
the form of  combined plots  of  $A_{FB}$ versus $A_{LR}$ and
 $A^{Pol}_{FB}$ versus $A_{LR}$,
 at LHC, for $M_{Z'}=1$ TeV.
Note that for $M_{Z'}=2$ TeV the figures are very similar. 
The error bars correspond to the
statistical error obtained with 1000 $e^+e^-+1000\ \mu^+\mu^-$ events
(polarization of the beam is assumed to be 100\%). 

\begin{figure}[h]
\centerline{\psfig{file={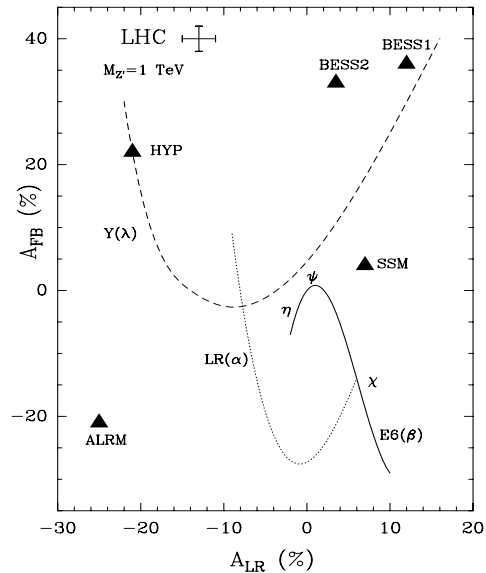},width=8truecm,height=10truecm}}
\vspace{-1.3cm}
\caption{$A_{FB}$ versus $A_{LR}$ according to the various
models.}
\end{figure}

\begin{figure}[h]
\centerline{\psfig{file={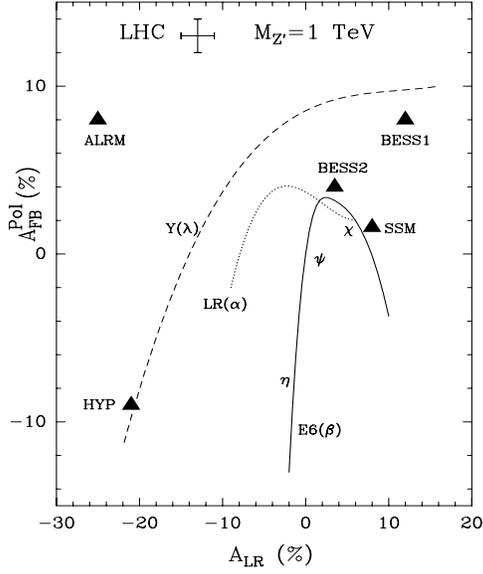},width=8truecm,height=10truecm}}
\vspace{-1.3cm}
\caption{ $A_{FB}^{Pol}$ versus $A_{LR}$ according to the various
models.}
\end{figure}

As can be seen from Figs.2-3, it is now very easy to
separate the various models, appart from very particular points
where some accidental crossing occurs due to the equality of the
couplings.

\subsection{Hadronic channels}

The example above shows that beam polarization is useful
to analyse some new physics effects but not to discover them. It is
a peculiarity of the leptonic channels. For hadronic channels, like jet production
we recover similar results as in the RHIC case.
Indeed, for $qq$-CI one expect a limit of the order of
$\Lambda\simeq 25\, TeV$ from the unpolarized jet cross section 
with $L=30\, fb^{-1}$
and $\sqrt{s}=14\, TeV$ \cite{Atlas}. With the same $L$ and $\sqrt{s}$ values
but with polarized protons, we obtain $\Lambda\simeq 50\, TeV$
from PV spin asymmetries.
For leptophobic $Z'$ similar results are obtained : the unpolarized
jet cross section is sensitive to masses up to $2\, TeV$, but 
one can reach $3.5\, TeV$ thanks to PV spin asymmetries.\\
Consequently, we stress again that polarized beams are well suited to
discover a new weak force in the hadronic sector.

\end{document}